\documentstyle[12pt]{article}

\thicklines                         
\def\a{\alpha}
\def\b{\beta}

\def\d{\delta}
\def\f{\phi}                    
\def\g{\gamma}

\def\k{\kappa}

\def\m{\mu}
\def\n{\nu}

\def\s{\sigma}                  

\def\D{\Delta}

\def\ca{{\cal A}}

\def\cs{{\cal S}}

\def\bo{\raisebox{-.4ex}{\large$\Box$}}                 
\def\cbo{{\,\raise-.15ex\Sc [\,}}                       
 
\def\svev#1{\left\langle #1\right\rangle}       

\def\ddt#1{{\buildrel {\hbox{\LARGE .\kern-2pt.}} \over {#1}}}

\def\beq{\begin{equation}}
\def\eeq{\end{equation}}
\def\bqry{\begin{eqnarray}}
\def\eqry{\end{eqnarray}}

\def\beqn#1{ \renewcommand{\theequation}{#1} 
             \begin{eqnarray} }
\def\eeqn{ \renewcommand{\theequation}{\arabic{equation}}
           \end{eqnarray} }

\def\beqr#1{ \setcounter{equation}{#1} 
             \begin{eqnarray} }

\def\eeqr{\end{eqnarray}}
\def\NON{\nonumber\\}
\def\beqrabc#1{ \setcounter{equation}{0}
                \renewcommand{\theequation}{#1\alph{equation}} 
                \begin{eqnarray} }
\def\beqrn#1#2{ \setcounter{equation}{#2}
                \renewcommand{\theequation}{#1.\arabic{equation}} 
                \begin{eqnarray} }
\def\seeq#1{eq.~(\ref{#1})}

\def\seeqs#1{eqs.~(\ref{#1})}

\def\seneq#1{~(\ref{#1})}
\def\rf{ref.~\cite}

\def\JMP#1{Jour. Math. Phys. {\bf #1}}
\def\NPB#1{Nucl. Phys. {\bf B#1}}
\def\NPBP#1{Nucl. Phys. (Proc. Suppl.) {\bf B#1}}
\def\PLB#1{Phys. Lett. {\bf B#1}}

\def\PRL#1{Phys. Rev. Lett. {\bf #1}}

\def\sstyle{\scriptstyle}

\def\frac#1#2{ {\sstyle {#1\over #2} } }

\def\tr{{\rm tr}\,}

\def\hc{{\rm h.c.\,}}

\def\Re{{\rm Re\,}}

\renewcommand{\thefootnote}{\fnsymbol{footnote}}

\def\tk{\tilde\k}

\begin{document}
\hyphenation{fer-mio-nic per-tur-ba-tive}

\noindent August 1996 \hfill TAUP--2361--96

\noindent revised December 1996 \hfill UTCCP-P-11

\hfill Wash. U. HEP/96-60

\hfill hep-lat/9608116
\par
\begin{center}
\vspace{15mm}
{\large\bf A Gauge-Fixing Action for Lattice Gauge Theories}
\\[10mm]
Maarten F.L.\ Golterman$^a$\footnote{permanent address: Dept. of Physics,%
~Washington University,%
~St. Louis, MO 63130, USA} and Yigal Shamir$^b$
\\[10mm]
$^a\!\!$ {\it Center for Computational Physics\\
University of Tsukuba\\
Tsukuba 305, Japan}\\
email: maarten@aapje.wustl.edu
\\[5mm]
$^b\!\!$ {\it School of Physics and Astronomy\\
Beverly and Raymond Sackler Faculty of Exact Sciences\\
Tel-Aviv University, Ramat Aviv 69978, Israel}\\
email: ftshamir@wicc.weizmann.ac.il
\\[15mm]
{ABSTRACT}
\\[2mm]
\end{center}

\begin{quotation}

  We present a lattice gauge-fixing action $S_{gf}$ with the following 
properties: (a) $S_{gf}$ is proportional to the trace of 
$(\sum_\m \partial_\m A_\m)^2$, 
plus irrelevant terms of dimension six and higher; 
(b) $S_{gf}$ has a unique absolute minimum at $U_{x,\m}=I$. 
Noting that the gauge-fixed action is not BRST invariant on the
lattice, we discuss some important aspects of the phase diagram.

\end{quotation}

\setcounter{footnote}{0}
\renewcommand{\thefootnote}{\arabic{footnote}}

\newpage

\noindent
{\it 1.}  Gauge theories are presently being investigated in two ways:
in the continuum (mostly) using perturbation theory, and on the lattice, 
which is
the method of choice for nonperturbative calculations.  In the continuum,
the classical action needs to be changed in order to define the quantum
theory; this is done by gauge fixing, which takes care of the otherwise
ill-defined integration over gauge orbits. In the usual lattice approach 
the integration over the gauge group is well defined due to the compactness 
of the gauge group, and gauge fixing is not necessary. However, it is
interesting to ask whether the continuum approach can be successfully
implemented nonperturbatively, {\it i.e.} on the lattice.

  A concrete proposal on how to do this was put forward in \rf{sml}  
(henceforth called $I$) in an attempt to define lattice chiral gauge
theories.  Because of the fact that gauge invariance is broken when one
regulates chiral gauge theories, it may in fact be necessary to gauge-fix 
the theory on the lattice, as has been suggested some years ago
in \rf{roma}.  (See also \rf{rev} for a recent review.)
The central observation is that, due to the lack of gauge invariance,
the longitudinal component of the gauge field couples to the fermions. 
It therefore becomes important to have a good control
over the dynamics of the longitudinal degree of freedom. A natural way to
achieve this is via gauge fixing, which can provide a kinetic term
(as well as possible interaction terms) for the longitudinal component.

  In $I$ a lattice gauge-fixing action was proposed which in the continuum 
limit leads to a nonlinear gauge-fixing action of the form 
$(\partial \!\cdot\! A + g A^2)^2$. 
This gauge is not suitable for $SU(N)$ theories, and
is also less familiar.  In this paper, we propose a lattice version of the
usual covariant gauge-fixing term $(\partial \!\cdot\! A)^2$. In addition
to presenting the form of the lattice action, we show that it has a single
absolute minimum at $U_{x,\mu}=I$ ($U_{x,\mu}$ is the compact lattice
gauge field), and we discuss some aspects of the phase diagram.  

  A central feature of our gauge-fixing approach is that
the gauge-fixed lattice action will {\it not} be
invariant under BRST transformations. This is important, because it was
shown that, nonperturbatively, a BRST-invariant partition function,
as well as expectation values of BRST-invariant operators, 
vanish as a consequence of the existence of lattice Gribov
copies~\cite{brst}. (If one is interested in
perturbation theory only, BRST symmetry can actually be
maintained on the lattice, see for instance \rf{llh} and references therein.) 
This implies that the 
vector boson mass term (along with other BRST symmetry violating
relevant and marginal operators) {\it must be tuned to zero by
hand}~\cite{roma}. The continuum limit is therefore characterized
by a vanishing second derivative at the minimum of the potential for the
{\it vector field}. This condition defines the boundary between a conventional
Higgs or Higgs-confinement phase, and a new phase (denoted FMD) 
which is characterized by the condensation of a vector field.
Below, the transition between the rotationally invariant
phase and the FMD phase will be denoted as ``the FMD transition.''

A condensate which breaks rotational symmetry appears strange at
first.  However, perturbation theory with a regulator that is not
gauge invariant already points at such a phenomenon.  Because of the
absence of gauge invariance of the regulated theory, a gauge-field
mass counterterm will be needed, with a parameter that needs to be tuned.
In the lattice version of the theory, one can envisage choosing this
parameter ``too small.'' The resulting negative value for the renormalized 
squared gauge-boson mass suggests that spontaneous symmetry breakdown occurs. 

\vspace{3ex}
\noindent
{\it 2.}  
In this section,
we will construct a gauge-fixed lattice action $S(U_{x,\m})$ which is 
invariant
under {\it global} gauge transformations in the gauge group, $G$ 
(but of course not 
under local transformations in this group). We will then show in the next
section that this action has the desired properties.

It will be convenient to separate out the
longitudinal degree of freedom by
introducing an additional group-valued scalar field $\f_x$,
and replacing $U_{x,\m}\to
\f_x^\dagger U_{x,\m}\f_{x+\hat\m}$. (This is a standard trick~\cite{phi}
that leads to a new, mathematically equivalent, formulation of the theory.)
The result is a Higher Derivative (HD)
action $S(\f_x^\dagger U_{x,\m}\f_{x+\hat\m})$,
and the symmetry group is enlarged to $G_{\rm local}\times G_{\rm
global}$:
\beq
U_{x,\m}\to h_x U_{x,\m}h_{x+\hat\m}^\dagger,\ \ \ \f_x\to h_x\f_x
g^\dagger,
\label{symm}
\eeq
where $g\in G_{\rm global}$ represents the original global symmetry.
One can regain the original formulation by choosing $\f_x=I$, which
amounts to gauge-fixing the enlarged, unphysical local symmetry.  
The only symmetry
present in that case is the global symmetry
\beq
U_{x,\m}\to g U_{x,\m}g^\dagger, 
\label{gsymm}
\eeq
(setting $h_x=g$ in \seeq{symm}).
This is the symmetry that we require to enlarge to a
local symmetry in the continuum limit.  
The field $\f_x$ 
explicitly represents the unphysical gauge degrees of freedom, which
couple to the transversal degrees of freedom because the lattice action $S(U)$
is not invariant under local $g$-transformations~\cite{phi}. 

One starts with a simple model that 
gives rise to the FMD transition described above. The action
(see Sect.~3.a and 4.b of $I$), which borrows from previous work on 
higher derivative actions~\cite{jkl}, is given by 
\beq
  S_H = \tr \sum \left(
         -\k\, \f^\dagger\bo(U)\f + \tk\, \f^\dagger \bo^2(U) \f \right)\,,
\label{sh}
\eeq
where 
\beq
  \bo_{xy}(U) = \sum_\m ( \d_{x+\hat\m,y} U_{x,\m} 
  + \d_{x-\hat\m,y} U^\dagger_{y,\m}) - 8 \d_{x,y} \,,
\eeq
is the standard nearest-neighbor covariant laplacian, and the lattice
spacing $a$ is set equal to one. 
For the gauge field we will assume the standard plaquette action.
(Because $\f_x$ is unitary, 
$S_H$ is in fact a function of $\f_x^\dagger U_{x,\m}\f_{x+\hat\m}$,
in agreement with the discussion above.)

Since we are interested in taking the gauge coupling $g_0$ to be
small, it is relevant to consider the {\it reduced} model,
which is obtained by turning off the gauge field in \seeq{sh}
altogether. The symmetry of the reduced model is 
$G_{\rm global} \times G_{\rm global}$, which corresponds to setting
$h_x=h$ in \seeq{symm}.
Here we will focus on the region of the phase diagram with
$\tk\to \infty$. 
The idea is that in the model with no gauge field $\tk\to\infty$ 
is a zero temperature limit where $v\equiv\langle\f\rangle\to I$.
(Note that $\langle\f\rangle$ is well-defined in the reduced model.) 
This breaks the symmetry 
$G_{\rm global} \times G_{\rm global} \to G_{\rm global}$ (with $h=g$ in
\seeq{symm}) consistent with the symmetry of the
formulation with the scalar field fixed to $\f_x=I$ ({\it cf.} \seeq{gsymm}). 
Had we chosen $\tk$ such that we were
close to a phase transition where $v$ becomes small, $\f_x$ would
develop a radial mode dynamically, and new, undesired excitations would
be present in the continuum limit.
All this implies that the continuum limit must be taken well inside
some {\it broken phase} of the reduced model. 
(We note that -- in the case of chiral gauge theories --
an alternative explanation as
to why the continuum limit must be taken in a broken phase is
provided by a generalized No-Go theorem~\cite{ys}, which asserts (modulo
some delicate loopholes) that the fermion spectrum in any symmetric
phase is vector-like.)  

The continuum limit, then, will be defined by approaching the
gaussian critical point $g_0=1/\tk=0$ on the FMD phase
boundary. Since $v\to I$ for $\tk\to\infty$, setting $\f_x=I$ (which amounts
to fixing the gauge for $G_{\rm local}$ in \seeq{symm}) provides the 
starting point for a valid 
perturbative expansion around this critical point.  (For some additional
discussion of the HD version of our action, we refer to the
conclusion.)
The usual weak coupling 
expansion $U_{x,\m} = \exp(ig_0 A_{x+{\hat\m\over 2},\m})$ then gives
\bqry
  \left. S_H  \right|_{\f_x=I} & = & 
  \k\, g_0^2\, \tr \sum_\m \Big( A^2_\m - {g_0^2\over 12} A^4_\m + \cdots 
\Big)
\NON
  & & +\, \tk\, g_0^2\, \tr \left( \Big( \sum_\m \D^-_\m A_\m \Big)^2
  + g_0^2 \Big( \sum_\m A^2_\m \Big)^2 + \cdots \right) \,,
\label{shexp}
\eqry
(where we have suppressed coordinates summations). Note in particular
the presence of the Lorentz symmetry violating term $\sum_\m A^4_\m$.
The dots in \seeq{shexp} stand for irrelevant operators.
$\D^-_\m$ is the backward lattice derivative, defined for an 
arbitrary function $f_x$ as $\D^-_\m\, f_x = f_x - f_{x-\hat\m}$. 

  As can be seen from \seeq{shexp}, the $\tk-$term in the action \seeq{sh}
leads to a kinetic term for the longitudinal part of the vector field.
Motivated by this observation, we set 
\beq
  \tk\, g_0^2 \equiv (2\a_0)^{-1} \,,
\label{a0}
\eeq 
where $\a_0$ is assumed to be a parameter of order one. 
Thanks to the presence of kinetic terms for {\it all}
polarizations, the vector lagrangian that governs the critical region
is manifestly renormalizable.
The $\k-$term in \seeq{shexp} is seen to lead to a mass term for
the vector field.  As shown in more detail in $I$ and below,
at tree level the FMD transition occurs at $\k=0$, with a non-zero
vector condensate for $\k<0$. 

  Now, we are interested in recovering a Yang--Mills theory in the 
continuum limit. To this end a renormalizable, but otherwise arbitrary, 
vector lagrangian will not suffice. What we need first is that
the tree-level lagrangian will agree with the continuum lagrangian
of a gauge theory, when the latter is quantized in a renormalizable 
gauge. Moreover, the lattice-regularized perturbation expansion
explicitly breaks the gauge invariance of the 
target continuum theory.
Therefore, the BRST identities must be enforced order by order in
perturbation theory (the issue of nonperturbative tuning will not
be addressed in this letter). A major role is played by the BRST identity
that requires the renormalized vector-boson mass to vanish;
this defines the location of the FMD transition.
This fact is at the heart of our 
approach. Thus, $\k$ is tuned to $\k_{c.l.}$ where in perturbation theory
the latter is given as a power series 
\beq
  \k_{c.l.}(g_0,\a_0) = \sum_{n\ge 1} c_n(\a_0)\, g_0^{2(n-1)} \,.
\label{kcl}
\eeq
Note the absence of an $O(g_0^{-2})$ term on the 
righthand side of \seeq{kcl},
in accordance with the requirement that the tree-level vector-boson mass
vanish.

In order to obtain a gauge-fixed continuum action,
the marginal terms on the second line of 
\seeq{shexp} should be of the form 
\beq
  {1\over 2 \a_0}\, (\mbox{gauge condition})^2 \,,
\label{gcond}
\eeq
for some gauge condition. Hence we need an additional term 
that, without spoiling the phase diagram, will bring the
marginal gauge-symmetry violating terms in the vector lagrangian into the
form\seneq{gcond}.

  Clearly, one has two options. The new marginal term can
be chosen to cancel the quartic term on the second row of \seeq{shexp}.
In this case only the bilinear term will remain, which corresponds to the
standard covariant gauge $\partial \!\cdot\! A =0$. Alternatively, the new
marginal term can be a mixed term proportional to 
$(\sum_\m \partial_\m A_\m)(\sum_\n A^2_\n)$. In this case one recovers
the nonlinear gauge $(\partial \!\cdot\! A + g A^2) =0$,
which was used in $I$. The main disadvantage
of this choice is that this nonlinear gauge condition is consistent only
for $U(1)$ or $SU(N)\times U(1)$. 

  We will now present a new HD action $S_{HD}$, 
with corresponding gauge-fixing action
\beq
  S_{gf}(U) \equiv S_{HD}(\f,U)  \Big|_{\f_x=I} \,,
\label{sgf}
\eeq
that enjoys the following properties:

\begin{itemize}

\item $S_{gf}$ admits the  expansion
$S_{gf} = {1\over 2 \a_0}\, \tr (\sum_\m \partial_\m A_\m)^2 
+ \mbox{\rm irrelevant terms}$.

\item $S_{gf}$ has a unique absolute minimum at $U_{x,\m}=I$. 

\item The important features of the phase diagram are unchanged. 

\end{itemize}

\noindent The new HD action is given by
\beq
  S_{HD} = {1\over 2 \a_0 g_0^2}\, \tr \sum \left(
           \f^\dagger \bo^2(U) \f - B^2 \right) \,,
\label{shd}
\eeq
\beq
  B_x = \sum_\m \left( {V_{x-\hat\m,\m} + V_{x,\m} \over 2} \right)^2 \,,
\label{B}
\eeq
\beq
   V_{x,\m} = {1\over 2i}
   \left(\f^\dagger_x U_{x,\m} \f_{x+\hat\m} - \hc \right) \,.
\label{V}
\eeq
(Note that the above definition of $V_\m$ leaves out a $g_0^{-1}$ factor
present in the corresponding definition in $I$.) From the point of view
of the HD model, 
$V_\m$ is a gauge-invariant vector field whose
expectation value can be used as an order parameter for the FMD phase.
Below we will use the alternative formulation where the 
field $\f_x$ is eliminated (see \seeq{sgf}). From the point of view of the 
weak coupling expansion,
one has $V_\m \to g_0 A_\m + O(g_0^3)$. Thus, the reader can easily check
that the unwanted $(\sum_\m A^2_\m)^2$ term in \seeq{shexp} is canceled
by the new term.

\vspace{3ex}
\noindent
{\it 3.}  Let us now discuss the properties of $S_{gf}$ in more detail.
Introducing 
\beq
  C_x = - \sum_y \bo_{xy}(U) \,,
\label{C}
\eeq
one can write 
\beq
  S_{gf} = {1\over 2 \a_0 g_0^2} \sum_x \cs_x \,,
\eeq
where
\beq
  \cs_x = \tr \left( C_x^\dagger C_x - B^2_x \right)\,.
\label{sx}
\eeq
Decomposing $C_x$ into its hermitian and anti-hermitian parts and using
cyclicity of the trace, one has $\cs_x=\cs_x^{(1)}+\cs_x^{(2)}$
where
\bqry
  \cs_x^{(1)} & = & \tr \left( {C_x^\dagger - C_x \over 2i} \right)^2 \,,
\label{sx1} \\
  \cs_x^{(2)} & = & \tr \left( {C_x^\dagger + C_x \over 2} + B_x \right) 
                        \left( {C_x^\dagger + C_x \over 2} - B_x \right) \,.
\label{sx2}
\eqry
Substituting \seeq{C} into \seeq{sx1} leads to
\beq
  \cs_x^{(1)} = \tr \left( \sum_\m \D^-_\m V_{x,\m} \right)^2 \,.
\label{marg}
\eeq
The expression inside the brackets is recognized as a lattice
transcription of the continuum $\sum_\m \partial_\m A_\m$. Thus, $\cs_x^{(1)}$
provides the desired longitudinal kinetic term, up to irrelevant operators.
These irrelevant terms are innocuous as long as one stays near the classical
vacuum $U_{x,\m}=I$. However, $\cs_x^{(1)}$ is known to have a host
of other zeros along the trivial orbit.
These minima are {\it lattice artifact Gribov copies}
of the classical vacuum. As argued in $I$, if one were to use only 
$\cs_x^{(1)}$
in the gauge-fixing action, one would end up with a 
phase diagram that differs qualitatively from the desired one.

  (We note in passing that the zeros of $\sum_\m \D^-_\m V_{x,\m}$ 
correspond to extrema of the functional $\Re \tr \sum_{x,\m} U_{x,\m}$
on a given gauge orbit. The choice of $\cs_x^{(1)}$ as the gauge-fixing 
action would assign equal probability to {\it all} extrema of this
functional. On the other hand, when one speaks about the lattice Landau gauge,
one usually refers to picking the global maximum of that functional.
For a manifestly gauge-invariant theory, such as lattice QCD, 
this is believed to be a valid gauge-fixing procedure. 
But here we want to be able to use an action to generate 
configurations which 
is not gauge invariant. Hence, the lattice Landau gauge method
and other nonlocal methods such as the use of the laplacian gauge~\cite{vink} 
introduce a genuine nonlocality.
It is very difficult to check whether this nonlocality disappears
in the continuum limit. If it does not, this may entail some inconsistency
in the analytic continuation back to Minkowski space. 
Our method avoids all these difficulties because
locality is manifestly preserved.)

  The role of $\cs_x^{(2)}$ is to cure the above problem. As we will now show,
$\cs_x^{(2)}$ contains only irrelevant operators, and its unique
absolute minimum is at $U_{x,\m}=I$. This validates weak coupling perturbation
theory, and lattice artifact Gribov copies are 
suppressed by $\cs_x^{(2)}\sim {\rm constant}/(\a_0g^2_0)$.
$\cs_x^{(2)}$ breaks BRST invariance explicitly, since 
$\cs_x^{(1)}+\cs_x^{(2)}$ cannot be written as the square of a local
gauge-fixing condition on the lattice.  

  Our aim is to prove that $\cs_x^{(2)}$ is nonnegative, 
and that it vanishes only for $U_{x,\m}=I$.
The trace of the product of two positive matrices is positive, and the
positivity of $(C_x^\dagger + C_x)/2 + B_x$ is obvious.
Consequently, the positivity of $\cs_x^{(2)}$
will follow once we show that $(C_x^\dagger + C_x)/2 - B_x$ is a positive 
matrix too. It is a straightforward exercise to check that
\beq
  (C_x^\dagger + C_x)/2 - B_x = \sum_\m \left( 
                                D_{x,\m}^{(1)} + D_{x,\m}^{(2)} \right) \,,
\label{D}
\eeq
where
\bqry
  D_{x,\m}^{(1)} & = & \left( I - {1\over 4} \Big(
  U_{x,\m} + U_{x-\hat\m,\m} + \hc \Big) \right)^2 \,,
\label{D1} \\
  D_{x,\m}^{(2)} & = & {1\over 2} I - {1\over 8} \Big( 
  U_{x,\m}^\dagger U_{x-\hat\m,\m} + U_{x,\m} U_{x-\hat\m,\m}^\dagger + \hc
  \Big) \,.
\label{D2}
\eqry
The positivity of $D_{x,\m}^{(1)}$ is manifest, whereas the 
positivity of $D_{x,\m}^{(2)}$ follows from the unitarity of the 
link variables.

  We next show that $\cs_x^{(2)}=0$ {\it iff} $U_{x,\m}=I$. For the abelian
case this statement is trivial to check. In the nonabelian case,
the condition $\cs_x^{(2)}=0$ requires that there exists an orthogonal basis, 
such that each basis vector is a zero eigenvector of 
$(C_x^\dagger + C_x)/2 + B_x$ and/or $(C_x^\dagger + C_x)/2 - B_x$. 
Now, a zero eigenvector of the sum of
two positive matrices must be a common zero eigenvector
($v^\dagger Mv=0\Leftrightarrow Mv=0$ for any positive hermitian matrix $M$).
Note that $(C_x^\dagger + C_x)/2 + B_x$ is explicitly the sum of
two positive matrices and, in view of \seeq{D}, a similar statement applies 
to $(C_x^\dagger + C_x)/2 - B_x$. Therefore, each of the above basis vectors 
must in particular be a zero eigenvector of $C_x^\dagger + C_x$
and/or $\sum_\m D_{x,\m}^{(1)}$. It is easy to check that the
zero eigenvectors of $C_x^\dagger + C_x$ and $\sum_\m D_{x,\m}^{(1)}$
are in fact common. They occur {\it iff} for all $\m$, 
$U_{x,\m}$ and $U_{x-\hat\m,\m}$
have a common submatrix equal to the identity. Thus, the condition
$\cs_x^{(2)}=0$ requires that both $C_x^\dagger + C_x$ 
and $\sum_\m D_{x,\m}^{(1)}$ be zero simultaneously, which is true
{\it iff} $U_{x,\m}=U_{x-\hat\m,\m}=I$. The proof is valid for unitary
and orthogonal groups. It can probably be generalized to any compact group.

  Lastly, we wish to check that $\cs_x^{(2)}$ contains only irrelevant
operators. One has the following expansion
\beq
  (C_x^\dagger + C_x)/2 + B_x = 2 g_0^2 \sum_\m A^2_\m + \cdots,
\label{expndC}
\eeq
where only the lowest dimensional operator is shown. Similarly,
\bqry
  D_{x,\m}^{(1)} & = & {g_0^4\over 4} A^4_\m + \cdots, 
\label{expndD1} \\
  D_{x,\m}^{(2)} & = & {g_0^2\over 4} \Big(\D^-_\m A_\m\Big)^2 + \cdots. 
\label{expndD2} 
\eqry
 From these expansions it follows that $\cs_x^{(2)}$ only contains operators
of dimension six and higher.

  We now digress momentarily to close a gap in the formulation of the
nonlinear gauge presented in $I$. While the minimum
of the classical potential was shown to be
$A_\m=0$, it was not established that $A_\m=0$ remains the absolute minimum
when $A_\m$ is allowed not to be constant. A lattice action that features the
same properties as \seeq{shd}, except that the marginal terms 
correspond to the continuum gauge-fixing action 
$(\partial \!\cdot\! A + g A^2)^2$, is given by 
\beq
S_{HD}^{n.l.} = {1\over 2 \a_0 g_0^2}\, \tr \sum \left(
           \f^\dagger \bo^2(U) \f + 2 B \sum_\m \D^-_\m V_\m \right) \,.
\label{hdnonlin}
\eeq
The corresponding gauge-fixing action density is
\beq
 \cs_x^{n.l.} = \tr \left( \sum_\m \D^-_\m V_{x,\m} + B_x \right)^2
 + \cs_x^{(2)} \,,
\label{sxnonlin}
\eeq
where $\cs_x^{(2)}$ is the same as in the linear case ({\it cf.} \seeq{sx2}).
Thus, in both cases the same irrelevant operator
$\cs_x^{(2)}$ is used to protect the uniqueness of the absolute minimum
at $U_{x,\m}=I$.

\vspace{3ex}
\noindent
{\it 4.}  We will now discuss some properties of the phase diagram 
of the theory defined by the new action \seeq{shd}. 
(We plan to present a more complete analysis elsewhere.)
For large $\tk$ (and small $g_0$)
one is in a broken phase, which could be an ordinary broken phase
or an FMD phase. As we will now see, 
the latter is characterized by a vectorial order parameter
that defines a preferred direction. (The large-$\tk$ rotationally-invariant 
region of the phase diagram is a Higgs  
or Higgs-confinement phase. With ``ordinary broken phase," we refer to the
large-$\tk$ properties of this phase.)  

In order to look for the FMD transition, we set $\phi_x=I$ and 
take $U_\m={\rm exp}(iA_\m)$ constant (assuming that translation invariance
is not broken), 
and we minimize the free energy with respect to $A_\m$. (Here we rescaled
$g_0 A_\m\to A_\m$.)
The task is simplified in the limit
$\tk\to\infty$, or equivalently $g_0\to 0$ ({\it cf.} \seeq{a0}),  
which is the region of the phase diagram where we want
to be anyway. In that case, the free energy is just the classical
potential for $A_\m$.  
As we will now show, $\k=0$ is the location
of the FMD transition classically, and a nonzero $\svev{A_\m}$
develops for
$\k<0$.

  In order to find the FMD transition for large $\tk$, we only have 
to keep the lowest
dimensional terms in the classical potential separately for the $\k$-
and $\tk$-terms. This leads to
\beq
  V_{cl} \approx \k\, \tr \sum_\m A_\m^2 +
  {\tk\over 2}\, \tr \Big( \sum_\m A_\m^2 \Big) \Big( \sum_\n A_\n^4 \Big) \,.
\label{vcl}
\eeq
(For $\k/\tk$ small, it is 
consistent to keep only the quadratic part of the $\k$-term.)
Minimizing this with respect to $A_\nu$, and taking the gauge group to
be $U(1)$, we obtain 
\beq
 \left[2\k + \tk \left( \sum_\m A_\m^4  +
  2 \Big( \sum_\m A_\m^2 \Big) A_\n^2 \right)\right]A_\nu = 0, 
  \ \ \ {\rm for\ all\ }\nu.
\label{min}
\eeq
Assuming $\k<0$, the minimum is found to be
\beq
  \svev{A_\m} = \pm \left( { |\k| \over 6\tk } \right)^{1\over 4}\,,
  \quad\quad \mbox{all } \m \,.
\label{VEVA}
\eeq
In the nonabelian case, one also has to take into account the contribution
from the plaquette term. For $SU(2)$ this leads to the
requirement that the $\svev{A_\m}$ commute. Up to a similarity
transformation, $\svev{A_\m}$ is equal to $\s_3$ times the righthand side
of \seeq{VEVA}. Note that the expectation
value points in one of the sixteen directions defined by the
lattice vectors $(\pm1,\pm1,\pm1,\pm1)$. This is not surprising since
the classical potential \seeq{vcl} is invariant only under the lattice
rotation group, but not under an arbitrary $O(4)$ rotation.
The vectorial expectation value leaves unbroken the subgroup of lattice
rotations in the hyperplane perpendicular to $\svev{A_\m}$.

  As follows from \seeq{VEVA}, the mean-field critical exponent now is 
$1/4$ rather than $1/2$ as found in $I$ 
in the nonlinear case.  This suggests that 
the new critical point is in fact a tricritical point in some larger
parameter space.  This is indeed the case.  Quantum corrections will 
require the addition of counterterms, and for constant $A_\mu$ the only
possible ones are $(\sum_\mu A_\mu^2)^2$ and $\sum_\mu A_\mu^4$
(we consider again the $U(1)$ case for simplicity).  So let
us consider a more general potential of the form
\beq
V = \k\, \sum_\m A_\m^2 +
\b\,(\sum_\mu A_\mu^2)^2 + \g\,\sum_\mu A_\mu^4 +
{\tk\over 2}\, \Big( \sum_\m A_\m^2 \Big) \Big( \sum_\n A_\n^4 \Big) \,.
\label{vgeneral}
\eeq
We assume $\tk>0$. Again, this approximation is self-consistent 
for $\tk$ large relative to the other couplings.
The minimization conditions now become
\beq
  \k + 2\b\,\sum_\mu A_\mu^2 + 2\g\,A_\nu^2+ 
  {\tk\over 2} \left( \sum_\m A_\m^4  +
  2 \Big( \sum_\m A_\m^2 \Big) A_\n^2 \right) = 0 \,,
\label{mingen}
\eeq
for all components of the vector field that do not vanish.  All    
nonvanishing components have to be equal (up to signs), and if we
set those all equal to $\ca$, assuming that $n\in\{1,2,3,4\}$ of them do not
vanish, we obtain
\beq
   {3\over 2} n\tk\,\ca^4+2(n\b+\g)\,\ca^2+\k=0\,.
\label{minred}
\eeq
The value $V_0$ of the potential at any extremum point can be written as
\beq
  V_0 = { n \ca^2 \over 3}
        \left( 2\k + (n\b+\g) \ca^2 \right) \,.
\label{V0}
\eeq
This expression is useful in studying the order of the transition.
Note that we are dealing here with a three parameter phase diagram,
spanned by $\b$, $\g$ and $\k$ ($\tk$ is large and can be scaled away
by absorbing it into $\ca$).
We will now show that for $\g + \min(\b,4\b) > 0$ there is a second order
transition, whereas for $\g + \min(\b,4\b) < 0$ the transition is first
order. Assume first $\g + \min(\b,4\b) > 0$. For any $\k>0$, the
lefthand side 
of \seeq{minred} is greater than zero, whereas for $\k=0$ the potential has a
quartic zero at the origin. Hence, $\k=0$ is a second order transition point.
Now assume $\g + \min(\b,4\b) < 0$. For $\k=0$,
\seeq{minred} has a solution $\ca^2>0$ for which $V_0<0$ at least for one $n$.
This implies that a first order transition has already occurred at some 
$\k>0$.
The first order surface joins the second order surface smoothly
at the tricritical line $\g + \min(\b,4\b)=0$, $\k=0$, 
separating a rotationally-invariant phase
($\ca=0$) from an FMD phase ($\ca\ne 0$).  Close to the tricritical line
the first order surface is $\k=(n\b+\g)^2/(2n\tk)$ where $n\b=\min(\b,4\b)$,
corresponding to $n=1$ for $\b>0$ and $n=4$ for $\b<0$. Away from the
tricritical line (and in the quadrant $\b<0$, $\g<0$) 
the value of $n$ may be different.

\vspace{3ex}
\noindent
{\it 5.}  In this letter we have addressed the question as to how a 
nonperturbative
({\it i.e.} lattice) definition can be given of a 
Lorentz gauge-fixed Yang--Mills
theory. In particular, we proposed a lattice version of the gauge-fixing
action that has a unique global minimum at $U_{x,\mu}=I$, 
and that has the correct classical continuum limit.  The
model can be studied in weak coupling perturbation theory. We also expect
that lattice artifact Gribov copies will 
be suppressed in the continuum limit.

  Because the lattice gauge-fixed action is not BRST invariant, the 
integration
over gauge orbits leads to nontrivial dynamics.  
We argued that a new second order phase transition
is expected between a Higgs or Higgs-confinement phase, and an FMD phase
which is characterized by the condensation of the vector field. 
As explained in the introduction, standard
perturbative arguments already indicate that
such a phase transition is {\it unavoidable} if one regulates a gauge
theory in a gauge {\it noninvariant} way.  A mass term for the gauge field
will generically appear, and will have to be subtracted in order to keep the
gauge field massless.  From the lattice point of view, this corresponds
to tuning to a continuous phase transition, and the desired continuum
theory corresponds to the critical theory (for $g_0\to 0$).  
(With a gauge-invariant
regulator, the gauge symmetry guarantees the theory to be at this 
critical point.) 

The details of this symmetry breaking, 
such as the critical exponent and the allowed directions of the condensate, 
depend on the gauge condition and the specific regulator employed.
As we showed, 
our lattice discretization of the Lorentz gauge-fixing action
corresponds to a tricritical point.  In practice, one should be
so close to the critical point that, within a given accuracy, the
nonphysical effects from being in either the Higgs or the FMD phase are
small enough.  This should be possible, since the phase transition is a
continuous one.

  There are two equivalent formulations for the lattice model we
presented in this letter.  The HD version makes the gauge degrees of
freedom explicit through the (unphysical)
group-valued scalar field $\f_x$, while the
gauge-fixed version is obtained by setting $\f_x=I$. This is {\it not} 
specific
to the lattice formulation of the model: the HD version of any gauge-fixed
continuum gauge theory can be defined in a similar way.  In perturbation
theory, of course, one usually does not introduce the scalar field.  On
the lattice, however, it is useful to do this, since that makes it
possible to first consider the {\it reduced} model -- 
the pure scalar theory obtained by setting $U_{x,\m}=I$ in the HD action.
The latter is more easily amenable to nonperturbative techniques
such as numerical simulation.  This brings up an interesting point: the 
quadratic part of the scalar action contains four-derivative terms, and
therefore raises the specter of infrared behavior divergences.  This
was discussed in some detail in $I$, where arguments were given that infrared 
divergences in fact do not arise. 
We expect this because of the intimate relation between
perturbation theory in the reduced model and in the full model,
and since in the full model one has standard IR behavior.
We intend to report on a more detailed 
investigation of this point in the near future.

  In order to complete the definition of the model, a Faddeev--Popov 
term (in the nonabelian case), and counterterms (of which the $\k$-term
and the $\b$- and $\g$-terms in \seeq{vgeneral}
already are examples) will have to be added.
The coefficients of the counterterms are calculable in perturbation theory.
Note that the divergent as well as the finite parts of the counterterms
are needed to recover the BRST identities. The $\f_x$ dependence can
be made explicit by replacing $g_0 A_\mu$ with $V_\mu$ defined in \seeq{V}.
Once the complete action is constructed, we may again study the phase diagram.
The Faddeev--Popov ghosts are of course crucial for unitarity of the
target continuum gauge theory.  However, their effects only come into
play at one loop (where the optical theorem would be violated
without ghosts), {\it i.e.} at order $g_0^2\sim 1/\tk$ (\seeq{a0}).
The interaction of the ghosts with $\f_x$ will therefore be suppressed
by $1/\tk$, and hence we expect that the ghosts will
not change the essential features 
of the FMD transition at large $\tk$. 
The effect of counterterms on the potential for
$A_\mu$ has already been discussed above. 

  First, however, a detailed investigation of the phase diagram(s) 
of the actions
given in \seeqs{sh} and\seneq{shd} with $U_{x,\mu}=I$ is in order.  
(In the reduced model, the FMD phase is characterized
by a nonzero momentum of the ferromagnetic groundstate~\cite{sml},
and the FMD transition is actually an FM-FMD transition
in the relevant part of the phase diagram.)
The nature of the FMD transition should be studied in more detail in order 
to find out whether
this approach to lattice gauge theories may lead to the same results as the 
standard (perturbative) continuum version and the usual gauge-invariant 
lattice
approach.  
Of course, after that many issues remain, such as the explicit
construction of ghost- and counterterms, and the inclusion of the full gauge 
field. Lattice artifact
Gribov copies should be investigated in more detail,
and then the problem of continuum Gribov copies should be addressed. 
If this program is successful, it may lead to a method for
constructing nonperturbative versions of gauge theories for which no
gauge-invariant formulation is known. Chiral gauge theories
constitute an example where gauge fixing appears to address the essential
problems that sofar have hampered attempts to define them on the lattice.

\bigskip
\noindent {\bf Acknowledgements}

We would like to thank Wolfgang Bock and Jim Hetrick for many discussions.
M.G.\ is supported by the US Department of Energy as an Outstanding Junior 
Investigator.
Y.S.\ is supported in part by the US-Israel Binational Science 
Foundation, and the Israel Academy of Science.



\begin{thebibliography}{99}

\bibitem{sml} Y.\ Shamir, hep-lat/9512019 (revised May 1996).

\bibitem{roma} A.\ Borelli, L.\ Maiani, G.-C.\ Rossi, 
R.\ Sisto and M.\ Testa, \PLB{221} (1989) 360; \NPB{333} (1990) 335;
L.\ Maiani, G.-C.\ Rossi and M.\ Testa, \PLB{292} (1992) 397.

\bibitem{rev} Y.\ Shamir, plenary talk at Lattice'95, Melbourne, Australia,
\NPBP{47} (1996) 212.

\bibitem{brst} B.\ Sharpe, \JMP{25} (1984) 3324;
H.\ Neuberger, \PLB{183} (1987) 337.

\bibitem{llh} M.\ L\"uscher, in {\it Fields, Strings and Critical Phenomena}
(Les Houches 1988), eds. E.\ Br\'ezin and 
J.\ Zinn-Justin (North-Holland, 1990).

\bibitem{phi} D.\ Foerster, H.B.\ Nielsen and M.\ Ninomiya,
\PLB{94} (1980) 135; J.\ Smit, \NPBP{4} (1988) 451;
S.\ Aoki, \PRL{60} (1988) 2109;
K.\ Funakubo and T.\ Kashiwa, \PRL{60} (1988) 2113.

\bibitem{jkl} K.\ Jansen, J.\ Kuti and C.\ Liu, \PLB{309} (1993) 119,
127;
\NPBP{30} (1993) 681, {\bf B34} (1994) 635, {\bf B42} (1995) 630;
J.\ Kuti, \NPBP{42} (1995) 113.

\bibitem{ys} Y. Shamir, \PRL{71} (1993) 2691; \NPBP{34} (1994) 590;
hep-lat/9307002.

\bibitem{vink} J.C.\ Vink, \PLB{321} (1994) 239. 

\end{thebibliography}
\end{document}